\documentclass[12pt,preprint,preprintnumbers,amsfonts,aps,nofootinbib]{article}

\usepackage{amssymb}
\usepackage{amsmath}
\usepackage{epsfig}

\usepackage{epsf}
\usepackage{graphicx}
\usepackage{amsfonts}
\usepackage{amssymb}

\usepackage{cite}

\makeatletter
\renewcommand\section{\@startsection {section}{1}{\z@}%
                                 {-3.5ex \@plus -1ex \@minus -.2ex}
                                   {2.3ex \@plus.2ex}%
                                   {\normalfont\large\bfseries}}
\renewcommand\subsection{\@startsection{subsection}{2}{\z@}%
                                   {-3.25ex\@plus -1ex \@minus -.2ex}%
                                     {1.5ex \@plus .2ex}%
                                     {\normalfont\bfseries}}
\renewcommand\subsubsection{\@startsection{subsubsection}{3}{\z@}%
                                   {-3.25ex\@plus -1ex \@minus -.2ex}%
                                     {1.5ex \@plus .2ex}%
                                     {\normalfont\itshape}}
\makeatother

\newcommand{\be}{\begin{equation}}
\newcommand{\ee}{\end{equation}}
\newcommand{\bea}{\begin{eqnarray}}
\newcommand{\eea}{\end{eqnarray}}
\newcommand{\barr}{\begin{array}}
\newcommand{\earr}{\end{array}}

\def\beq{\begin{equation}}
\def\eeq{\end{equation}}
\def\be{\begin{equation}}
\def\ee{\end{equation}}
\def\bea{\begin{eqnarray}}
\def\eea{\end{eqnarray}}

\DeclareRobustCommand{\SkipTocEntry}[4]{}

\begin{document}

\begin{titlepage}

\setcounter{page}{1} \baselineskip=15.5pt \thispagestyle{empty}

\begin{flushright}
SU-ITP-09/06 \\
SLAC-PUB-13536
\end{flushright}
\vfil

\begin{center}

{\Large \bf Trapped Inflation}
\\[0.7cm]
{\large Daniel Green$^{a}$, Bart Horn$^{a}$,  Leonardo Senatore$^{b,c}$ and Eva Silverstein$^{a}$}
\\[0.7cm]
{\normalsize {\sl $^{a}$ SLAC and Department of Physics, Stanford University,
Stanford CA 94305, USA}}\\

\vspace{.3cm}
{\normalsize { \sl $^{b}$ School of Natural Sciences, Institute for Advanced Study, \\Olden Lane, Princeton, NJ 08540, USA}}\\

\vspace{.3cm}
{\normalsize { \sl $^{c}$ Jefferson Physical Laboratory and Center for Astrophysics, Harvard University, Cambridge, MA 02138, USA}}\\



\end{center}

\vspace{.8cm}

\hrule \vspace{0.3cm}
{\small  \noindent \textbf{Abstract} \\[0.3cm]
\noindent We analyze a distinctive mechanism for inflation in which particle
production slows down a scalar field on a steep potential, and show
how it descends from angular moduli in string compactifications. The analysis of density perturbations -- taking into account the integrated effect of the produced particles and their quantum fluctuations -- requires
somewhat new techniques that we develop. We then
determine the conditions for this effect to produce sixty
e-foldings of inflation with the correct amplitude of density
perturbations at the Gaussian level, and show that these
requirements can be straightforwardly satisfied.  Finally, we estimate the amplitude of the non-Gaussianity in the power spectrum and find a significant equilateral contribution. }\vspace{0.5cm}  \hrule


%


%
%



\vfil
\begin{flushleft}
\today
\end{flushleft}

\end{titlepage}

\newpage
\tableofcontents
\newpage

\section{Introduction}

Inflation \cite{Inflation}\ is a very general framework for addressing the basic problems of primordial cosmology.  It requires a source of stress-energy which generates an extended period of accelerated expansion.  This can arise in many different ways even at the level of a single scalar inflaton, for which the space of inflationary models has been usefully organized by an effective field theory treatment \cite{Cheung:2007st}. The various mechanisms can be  distinguished in many cases via their distinct predictions for the CMB power spectrum and for relics such as cosmic strings that may be produced after inflation.
As well as being observationally accessible, inflationary theory is also sensitive to the ultraviolet completion of gravity, for which string theory is a promising candidate.

Traditional slow roll inflation requires a flat potential, which can be obtained naturally using approximate shift symmetries
or with modest fine-tuning.
Inflation, however, does not require a flat potential.  Rather, in general in single-field inflation \cite{Cheung:2007st,gensinglefield}\ the inflaton may self-interact in such a way as to slow itself down even on a steep potential as in e.g. \cite{Garriga:1999vw,Silverstein:2003hf,Alishahiha:2004eh,Chen:2004gc}.  It is interesting to examine such mechanisms further, to explore their novel dynamics and to better assess the level of fine-tuning required to obtain them from the point of view of both effective field theory and string theory.

In this work, we analyze a simple mechanism for inflation in which the inflaton $\phi$ rolls slowly down a steep potential by dumping its kinetic energy into the production of other particles $\chi_i$ (plus appropriate supersymmetric partners) to which it couples via interactions of the form
\beq\label{interaction} {1\over 2} g^2\sum_i(\phi-\phi_i)^2\chi_i^2\ . \eeq
As $\phi$ rolls past each point $\phi_i$, the corresponding $\chi_i$ particles become light and are produced with a number density that grows with increasing field velocity $\dot\phi$.  As it dumps energy into the produced particles, $\phi$ slows down; meanwhile the produced particles dilute because of the Hubble expansion. With sufficiently closely spaced points $\chi_i$ we will see that this yields inflation even on a potential that is too steep for slow-roll inflation.
This mechanism, trapped inflation, was originally suggested in \cite{Kofman:2004yc}\ based on the preheating mechanism developed by \cite{Kofman:1997yn}~\footnote{There are other interesting approaches using a gas of particles to slow the field evolution on a steep potential in order to inflate (see e.g. the recent review \cite{Berera:2008ar}\ and \cite{Berera:1999wt}) or to avoid the overshoot problem in small-field inflationary models (see e.g.~\cite{Kaloper:1991mq,Brustein:2004jp,Itzhaki:2007nk}).  The change in the CMB power spectrum from a single particle production event was also studied in \cite{Chung:1999ve}.}.  It can be usefully viewed \cite{Silverstein:2003hf}\ as a weak-coupling analogue of DBI inflation (or vice versa) in which the effects on $\phi$'s motion from the production of the $\chi$ fields dominates over their loop corrections to its effective action.

From the low energy point of view, although couplings of the general form (\ref{interaction}) are generic, the prospect of many closely spaced such points $\phi_i$ seems rather contrived.  However, we will see that just this structure descends from string compactifications in a rather simple way.   It arises in the same type of angular directions in field space that undergo {\it monodromy} from wrapped branes as studied recently in \cite{Silverstein:2008sg,McAllister:2008hb}.

%
%

In \cite{Silverstein:2008sg,McAllister:2008hb}, a single wrapped brane was considered.  A scalar $\phi$ rolls down the potential over a large distance $\Delta\phi\gg M_P$ corresponding to multiple circuits of an underlying circle around which the brane tension undergoes monodromy.  In this super-Planckian regime, the potential satisfies slow roll conditions as in chaotic inflation \cite{Linde:1983gd}\  (though with a distinctive power law behavior depending on the example).   In the same direction, at {\it sub}-Planckian field values $\phi\le M_P$, the potential is too steep for slow roll inflation.
However, in variants of these setups, because of the underlying small circle, the system periodically develops new light degrees of freedom as the inflaton rolls down the steep part of the potential.


The analysis of the perturbation spectrum in this class of models, including the integrated effects of the produced particles, has interesting novelties.  The number of produced particles fluctuates quantum mechanically, leading to a source term in the equation of motion for the perturbations of the inflaton.  A constant solution to the homogeneous mode equation develops parametrically before the mode stretches to the Hubble horizon,
as in previous examples of single field inflation in the presence of a low sound speed \cite{gensinglefield,ArkaniHamed:2003uz,Alishahiha:2004eh}.
Finally, as with other mechanisms such as \cite{Silverstein:2003hf,Alishahiha:2004eh}\ in which interactions slow the inflaton, a simple estimate reveals a correspondingly large non-Gaussian contribution to the perturbation spectrum in trapped inflation, which will be within the range tested by the upcoming Planck satellite \cite{Planck}\ according to preliminary estimates for its capacity to detect or constrain the three-point amplitude $f_{\rm NL}^{\rm equilateral}.$

While this work was in completion we received the interesting work \cite{kofmanetal}, which has some overlap with the present paper.

\section{Background Solution}
In this section we will find the background solutions and the
conditions for trapped inflation, without making use of the perturbation spectrum.  We
will discuss perturbations in the next section.  Getting the power
spectrum to match observation will further constrain our
parameters.

The idea of trapped inflation is that particle production will slow
the inflaton ($\phi$) enough to produce inflation on a potential
which would be too steep for slow-roll inflation. For this to work,
we will need a number of different fields to become massless at
regular intervals along the $\phi$ direction. A Langrangian
describing such a configuration can be written as
\beq \mathcal{L} = \frac{1}{2}\partial_{\mu}\phi
\partial^{\mu}\phi-V(\phi)+\frac{1}{2}\sum_{i}(\partial_{\mu}\chi_{i}
\partial^{\mu}\chi_{i}-g^2(\phi-\phi_i)^2\chi_{i}^2) + \ldots\ ,\eeq
where the $\dots$ represent the supersymmetric completion of these
terms, applicable in appropriate cases.  Softly broken supersymmetry helps to suppress Coleman-Weinberg corrections to the effective action arising from the loops of light $\chi$ particles.
As discussed in \cite{Silverstein:2003hf,Kofman:2004yc}, at weak coupling particle production dominates over quantum corrections to the effective action for colliding locally maximally supersymmetric branes.
Here the points $\phi_{i}$
are the points where $\chi_{i}$ become massless.  For simplicity, we
take these to be evenly spaced in $\phi$ with spacing $\Delta \equiv
\phi_{i+1}-\phi_i$. The coupling $g$ may be small.  If $\phi$ starts
rolling down the potential $V(\phi)$, whenever it crosses a point
$\phi_{i}$, $\chi_i$ particles are produced.  The expectation
value~\footnote{For the purposes of calculating the homogeneous background inflationary solution, the expectation value of $n_\chi$ is all we will need.  In calculating the perturbation spectrum in the next section, we will require its higher point correlation functions.} of the number density of
the $\chi_{i}$ particles produced is given by
\cite{Kofman:1997yn,Kofman:2004yc}
\beq\label{nchi} n_{\chi}(t) \simeq \frac{g^{\frac{3}{2}}}{(2\pi)^3}
(\dot{\phi}(t_{i}))^{\frac{3}{2}} \frac{a(t_{i})^3}{a(t)^3}\ , \eeq
where $t_{i}$ is defined by $\phi(t_{i})=\phi_i$ and the powers $a =
e^{\int^t H dt'}$ accounts for the dilution of particles due to the
expansion of the universe. The energy density of the $\chi$
particles is then given by $g|\phi-\phi_{i}| n_{\chi}$ following the
particle production event, i.e. once the system has passed back into
the adiabatic regime where $\frac{\dot{\omega}}{\omega^2} \ll 1$.
Because $m_{\chi} = g|\phi-\phi_{i}|$, the $\chi$ fields behave
adiabatically when \beq g \dot{\phi} \ll g^2 |\phi-\phi_{i}|^2\ . \eeq
Making the replacement $|\phi-\phi_{i}| \simeq \dot{\phi} \delta t$,
we can estimate the timescale on which the particle production happens: $\Delta t\sim (g\dot\phi)^{-1/2}$.
Requiring this timescale to be short compared to Hubble $\Delta t
\ll H^{-1}$ implies \beq \label{instant} H^2 \ll g \dot{\phi}\ .
\eeq\

\noindent The parametric dependence of eq. (\ref{nchi}) can be
understood by noticing that the particle are effectively massless at
production time, and are produced during a time of order $\Delta t$.
This explains why $n \sim 1/{\Delta t}^3$. On a longer timescale, Hubble dilution becomes important, and $n\propto a(t)^{-3}$.

The equations for motion for the homogeneous
background solution (including the energy density in $\chi$
particles) can be derived either from $D_{\mu} T^{\mu}_{\nu} =0$ or
by approximating $\chi\chi$ with $\langle \chi \chi \rangle$ in the
equations of motion for $\phi$ as explained in \cite{Kofman:1997yn}.
The $\phi$ equation of motion is
\beq \label{eom1} \ddot{\phi}+3H \dot{\phi}+V'(\phi)+\sum_{i}
\frac{g^{\frac{5}{2}}}{(2\pi)^3} (\dot{\phi}(t_{i}))^{\frac{3}{2}}
\frac{a(t_{i})^3}{a(t)^3} =0\ , \eeq
where $V' \equiv \frac{\partial V}{\partial \phi}$.  This sum over
particle production events will be difficult to deal with, so we
would like to replace it with an integral, giving us \beq
\label{eom2} \ddot{\phi}+3H \dot{\phi}+V'(\phi)+ \int^t
\frac{g^{\frac{5}{2}}}{\Delta(2\pi)^3}
(\dot{\phi}(t'))^{\frac{5}{2}} \frac{a(t')^3}{a(t)^3} dt' =0\ . \eeq
This is a good approximation to the sum when the variation of the
integrand is small between production events.  This is quantified by
the two conditions $\frac{H \Delta}{\dot{\phi}} \ll 1$ and
$\frac{\ddot{\phi} \Delta}{\dot{\phi}^2} \ll 1$.  Because of the
exponential suppression and the slow variation of the integrand, we
can replace the integral by \beq \int^t
\frac{g^{\frac{5}{2}}}{\Delta(2\pi)^3}
(\dot{\phi}(t'))^{\frac{5}{2}} \frac{a(t')^3}{a(t)^3} dt' \simeq
\frac{g^{\frac{5}{2}}}{3 H \Delta(2\pi)^3}
(\dot{\phi}(t))^{\frac{5}{2}}\ . \eeq This is a reasonable
approximation under the condition $\frac{\ddot{\phi}}{H \dot{\phi}}
\ll 1$.  If we assume slow roll and that the particle production is
the dominant mechanism for damping ($|\ddot{\phi}| \ll 3H |\dot{\phi}|
\ll V'$), then we can solve (\ref{eom2}) to get
\beq \label{sol}
\dot{\phi} \simeq -\frac{({3 H \Delta(2\pi)^3} V')^{\frac{2}{5}}}{g}\ .
\eeq

It is worth commenting on the limit $H\rightarrow 0$ of the above expression. In this case $\dot\phi$ goes to zero. This is due to the fact that in absence of dilution, the mass of the particles increases as $\phi$ moves after the time of particle-production. Therefore, $\phi$ loses energy even after the particles stop being produced. This explains why, in this $H\rightarrow 0$ limit, the solution is different from the $\dot\phi=$ const. that one would naively expect in the case of a linear potential. In the presence of a non-zero $H$, the growth in mass of the particles is compensated by their dilution, which allows for a steady solution $\dot\phi\simeq$ const. to exist.

Given the solution for $\phi$ in eq.~(\ref{sol}), we can find $H$ and the slow roll
parameters.  The usual Friedmann equation is
\beq \label{frw} 3 M_{P}^2 H^2 = \rho_{\phi}+\rho_{\chi} =
\frac{1}{2}\dot{\phi}^2+ V(\phi)+\sum_{i} g|\phi-\phi_{i}| n_{\chi}
\simeq V(\phi)\ . \eeq
%

We are assuming that the energy density is dominated by the
potential energy.  Using energy conservation $\dot{\rho} =
-3H(\rho+p)$ we get \beq 6 M_{P}^2 H \dot{H} = -3 H (\dot{\phi}^2+
\sum_{i} g|\phi-\phi_{i}| n_{\chi})\ , \eeq where we have used
$p_{\chi} \simeq 0$.  The generalized slow roll parameter $\epsilon
\equiv - \frac{\dot{H}}{H^2}$ is then given by \beq \epsilon =
\frac{ 3 (\dot{\phi}^2+ \sum_{i} g|\phi-\phi_{i}| n_{\chi}) }{2 V}\ .
\eeq As expected $\epsilon \ll 1$ is the statement that the energy
density is dominated by the potential.

We would like to use $\epsilon$ to constrain our parameters.  We
will assume that $\rho_{\chi} \gg \dot{\phi}^2$ so our condition
becomes
\beq V \gg \frac{3}{2} \int^{t} \frac{g^{\frac{5}{2}}}{\Delta (2
\pi)^3} |\phi(t)-\phi(t')| \dot{\phi}(t')^{\frac{5}{2}}
\frac{a(t')^3}{a(t)^3} dt'\ . \eeq
In order to constrain our parameters, we will make some estimates of
this integral.  Using $|\phi(t)-\phi(t')| \simeq \dot{\phi}(t-t')$
(given $|\ddot{\phi}| \ll |\dot{\phi}| H$), we can do the integral to
get
\beq V \gg \frac{3}{2} \frac{g^{\frac{5}{2}}}{9H^2 \Delta (2 \pi)^3}
\dot{\phi}^{\frac{7}{2}}\ . \eeq
Using (\ref{sol}), $V = 3M_{P}^2 H^2$ and dropping order one
factors~\footnote{In general, we will not keep track of all order one factors,
in part because our analysis of the integro-differential equation governing $\phi$ and its perturbations
will not be exact.}, we get
\beq \label{ineq1} (2\pi)^{6/5}\frac{V'^{\frac{7}{5}}
\Delta^{\frac{2}{5}}}{g M_{P}^2 H^{\frac{13}{5}}} \ll 1\ . \eeq

We are now in a position to massage some of our previous
inequalities to get conditions on individual parameters.  Using
(\ref{instant}), we can use our solution to get the inequality
\beq \label{ineq2} H^{\frac{8}{5}} \ll V'^{\frac{2}{5}}
\Delta^{\frac{2}{5}}(2\pi)^{6/5}\ . \eeq
This provides a lower bound on $\Delta$. The requirement that the
particle production events were frequent also gave us the inequality
$\Delta \ll |\dot{\phi}| H^{-1}$.  Using (\ref{sol}), this gives us
\beq \label{ineq3} \Delta^{\frac{3}{5}} \ll
\frac{V'^{\frac{2}{5}}}{g H^{\frac{3}{5}}}(2\pi)^{6/5}\ . \eeq
These two inequalities imply
\beq \label{ineq4} g H^3 \ll (2\pi)^3V'\ . \eeq

We can also use our constraints on $\ddot{\phi}$ to get analogues of
the slow-roll condition $\eta\ll 1$.  Recall that our solution
required $\ddot{\phi} \ll 3H \dot{\phi} \ll V'$. Taking a derivative
of (\ref{sol}) we get
\beq \ddot{\phi} = \frac{2}{5}(-\epsilon H
\dot{\phi}+\dot{\phi}^2\frac{V''}{V'})\ . \eeq
The first inequality, $\ddot{\phi} \ll 3H \dot{\phi}$ is trivially
satisfied for the first term, but the second gives us a new
condition \beq \label{eta} (2\pi)^{6/5}\frac{V''
\Delta^{\frac{2}{5}}}{g V'^{\frac{3}{5}} H^{\frac{3}{5}}} \ll 1\ .
\eeq The second inequality, $3H \dot{\phi} \ll V'$ also gives a
non-trivial condition
\beq \label{eta2} (2\pi)^{6/5}\frac{H^{\frac{7}{5}}
\Delta^{\frac{2}{5}}}{g V'^{\frac{3}{5}}} \ll 1\ . \eeq

There is another important requirement that we have ignored.
Inflation is required to last long enough to give at least 60
e-folds.
We will discuss this constraint in the context of an $m^2
\phi^2$ model, after we discuss perturbations.

\section{Perturbations}

\subsection{Gaussian Perturbations}

Determining the form of the curvature perturbation is a delicate
task.  Since the trapping is intrinsically a multifield effect, we have not developed a Langrangian description of our effective equation of motion for $\phi$ that one can consistently perturb. The strategy that we will use instead is to study the
perturbations using the equations of motion directly.

There are two approaches one could take.  The first is to use
constant $\phi$, `unitary', gauge and perturb in the metric.  This would seem to
have an obvious advantage, given that the particle production would
happen everywhere at the same time in this slicing.  Unfortunately,
solving the many constraint equations for the metric perturbation is
a complicated task. Similarly to what occurs in spontaneously broken gauge theories when one works in unitary gauge,  this would also be the gauge where the main physical degrees of freedom are most obscure. As is usually the case in inflation \cite{Cheung:2007st}, the matter scalar degree of freedom produces some scalar perturbations on the metric. These are not independent scalar degrees of freedom, but they are constrained variables. These perturbations of the metric are less important than the matter scalar excitations  (the scalar field $\phi$ and the $\chi$ particles in our case) for all the range of energies that we are interested in: from deep inside the horizon to freezeout. Thus, it is convenient to work in a gauge where the scalar field $\phi$ and the $\chi$ particles appear explicitly, so that one can neglect the metric perturbations. This leads us to the second possible approach to study the perturbations, which will be the one we take here, where we work with constant curvature slices. In reality, since metric perturbations are less  important, we will forget about them from the start, and we will work directly in an unperturbed quasi de Sitter universe.  We will therefore perturb our
equation of motion for $\phi$, taking into account the variance $\Delta n(x,t)$ in the number density of $\chi$ fields created.  After horizon exit, at the time of reheating, these are
converted to the curvature perturbation in the standard way~\cite{Bardeen}.

The equation of motion for $\phi(x,t)$ takes the form
\beq\label{eomphi}
\ddot{\phi}-\frac{\partial^2\phi}{a(t)^2} + 3H \dot{\phi}+V'(\phi)+\int^{t}
\frac{g^{\frac{5}{2}}}{\Delta(2\pi)^3}
(\dot{\phi}(t'))^{\frac{5}{2}} \frac{a(t')^3}{a(t)^3} dt'+g^2\sum_j(\chi_j^2-\langle\chi_j^2\rangle)(\phi-\phi_j)=0\ . \eeq
We have assumed as before that we can make the sum of sets of produced $\chi$ particles into an
integral (an approximation to be checked below), and we have included their quantum fluctuations in the last term.

In the Gaussian approximation, the last term in (\ref{eomphi}) is equivalent to the variance
in the number of produced $\chi$ particles:
\beq\label{variance} g^2 \sum_j(\chi_j^2-\langle\chi_j^2\rangle)(\phi-\phi_j) \simeq g \Delta n(x,t)\ . \eeq
This behaves as a source term in the equation for the inflaton perturbations.  This is somewhat analogous to the equation for perturbations discussed in \cite{warmperts}, and we can use some of the same techniques.
We will now perturb the $\phi$
field around the background solution as
\beq\label{fieldexpansion} \phi(x,t) =
\phi(t)+\varphi(x,t)\ . \eeq

When expanding our equation of motion in $\varphi$, we have to be careful to keep all the
contributing terms.  In particular, fluctuations of the inflaton
change the time when particle production occurs at different spatial
points.  This manifests itself as a fluctuation of $t'$, our
variable of integration, when we use the continuum approximation to
the sum over particle production events. We define $t'$ by $\phi_{i}
= \phi(x,t')$. Expanding in $\varphi$ and $t' \rightarrow
t'_{0}+\delta t'$, we find \beq \delta t' =
-\frac{\varphi}{\dot{\phi}}\ . \eeq We should think of the integral as
being over $t'_{0}$.  This implies that the upper limit of the
integral is also subject to the perturbation.  In particular, we can think
of the integral as being over all time with a step function
$\Theta(t-t'_{0}-\delta t')$.  This accounts for the fact that, on equal time slices, at
different spatial locations, a different number of $\chi$ fields
could have become massless and therefore been produced.  In these regions, a different number of
particles contribute to the sum, leading to a different region of
integration.

Putting these pieces all together we get the equation of motion for
the fluctuation
\begin{eqnarray}\label{expandedeom} && \ddot{\varphi}+\frac{k^2}{a^2}\varphi+3H
\dot{\varphi}+V'(\phi+\varphi)+\int^{t-\delta t'}
\frac{g^{\frac{5}{2}}}{\Delta(2\pi)^3} (\dot{\phi}(t'+\delta
t')+\dot{\varphi}(t'+\delta t'))^{\frac{5}{2}} \frac{a(t'+\delta
t')^3}{a(t)^3} dt'\nonumber \\
&& ~~~~~ = -g^2 \sum_j\left[(\chi_j^2-\langle\chi_j^2\rangle)(\phi+\varphi-\phi_j)\right]_k\ , \end{eqnarray}
where we have done a Fourier transform in
the spatial direction with $\frac{k}{a}$ being the physical
momentum.  For the Gaussian fluctuations, we will expand to linear
order in $\varphi$.  This gives an effective equation of motion
\beq
\label{gauss} \ddot{\varphi}+\frac{k^2}{a^2}\varphi + 3H
\dot{\varphi}+V''(\phi)\varphi+\varphi(t) \hat{m}^2 +
\int^{t} \hat{m}^2\left(\frac{5}{2} \dot{\varphi}(t')-
3H\varphi(t')\right)\frac{a(t')^3}{a(t)^3} dt'=-g \Delta n(k,t)\ , \eeq
where we have
defined $\hat{m}^2 \equiv \frac{ g^{\frac{5}{2}}}{ \Delta(2\pi)^3}
\dot{\phi}^{\frac{3}{2}}$ and used (\ref{variance}).  One can check that $V'' \ll
\hat{m}^2$ is the $\eta$-like condition (\ref{eta}), so we will
drop the $V''$ term.  We will see that this condition, {\it not} $V'' \ll H^2$, is sufficient to ensure that the spacetime is accelerating, and the modes freeze-out and produce curvature perturbations. This is very different than in the standard slow-roll case.

There are two types of contributions to the power spectrum -- those sourced by $\Delta n$, and those which would arise in the absence of the source.  We will find that the former dominates.  To begin, in order to analyze both these contributions, we require the homogeneous mode solutions to the above integro-differential equation.  This will
allow us to construct the Green's function required to determine the sourced perturbations.


To get some intuition for the behavior of the homogenous solutions, we will start by solving the equation
for constant $p \equiv k a^{-1}$.  This is a good approximation when
$\dot{p} p^{-2} \ll 1$ which holds until $p \simeq H$.  We will also
approximate $H$ and $\dot{\phi}$ as constant, which holds
to leading order in our generalized slow roll parameters.
There are three epochs of interest depending on the ratios $p/\hat
m$ and $p/H$:
%

\noindent{(I)}  $p\gg \hat{m}$:  The modes are approximately
Minkowskian, with both Hubble friction and particle production
effects negligible in their equations of motion; we start with the
pure positive frequency modes corresponding to the standard
Bunch-Davies vacuum.

\noindent{(II)} $H\ll p\ll \hat{m}$:  In this regime, a constant
solution to (\ref{gauss}) appears.  The mode solutions from region
I, evolved into region II, develop a term which is approximately
constant.
This contribution begins with a very small amplitude (which will
be determined in our exact solution below) 
but ultimately dominates over the other terms which become
damped exponentially in $Ht$.

\noindent{(III)} $p < H$:  In this regime, the curvature
perturbation $\zeta = \frac{H}{\dot\phi}\varphi$ becomes constant,
lying outside the Hubble horizon.

\smallskip

\noindent In particular, we will find that the modes actually
freeze-out well before reaching the Hubble horizon.  This is
somewhat analogous to the freeze-out of modes at the sound horizon
$c_s/H\ll 1/H$ in general single field models of inflation
\cite{Garriga:1999vw,ArkaniHamed:2003uz,Alishahiha:2004eh,Silverstein:2003hf}.

Now let us derive these features from a more detailed analysis of
(\ref{gauss}). For constant $p$, $\hat{m}$ and $H$, we can find
exact solutions to (\ref{gauss}) using the ansatz $\varphi(k,t)
\propto e^{\alpha t}$.  We can solve the equation trivially because
in our WKB regime of constant $p$, all terms are proportional to
$e^{\alpha t}$ with constant coefficients depending on $H$ and
$\alpha$.
In particular, using the ansatz and doing the integrals we find that
(\ref{gauss}) reduces to \beq \left(\alpha^2+3H\alpha + p^2+ \hat{m}^2
+ \hat{m}^2 \frac{ \frac{5}{2} \alpha - 3H}{3H+\alpha} \right)
e^{\alpha t} =0\ . \eeq This equation gives the mode solutions when
$\alpha \neq -3H$. Multiplying through by $3H+\alpha$, we get the
cubic equation \beq \label{mode} \alpha^3+6H\alpha^2+(9H^2 + p^2 +
\tilde{m}^2 )\alpha + 3H p^2 =0\ , \eeq where we have defined
$\tilde{m}^2 = \frac{7}{2} \hat{m}^{2}$.  It should be clear from
this equation that behavior of the perturbations will only be
different from the usual case if $\tilde{m}^2 \gg H^2$.  In this
model, this is always the case, as this condition is equivalent to
the slow roll condition $3H\dot{\phi} \ll V'$.

There are three analytic solutions to (\ref{mode}) since it is a
cubic.  To understand the behavior of the solution and impose
boundary conditions, it will be useful to expand these solutions
perturbatively in the different regimes discussed above. When $H^2
\ll p^2$, we can expand the modes around $H=0$, giving
\beq\label{solalphas} \alpha_{\pm} = \pm i \sqrt{p^2+\tilde{m}^2}-
\frac{3H}{2}\frac{2\tilde{m}^2+p^2}{ \tilde{m}^2+p^2}\ , ~~~~~~~
\alpha_{3} = - 3H \frac{p^2}{\tilde{m}^2+p^2}\ . \eeq
When $p^2 \gg \tilde{m}^2$ we can match onto the solutions in the
Bunch-Davies vacuum.  Specifically, we should use the mode
$\alpha_{+}$ with a normalization of $1/\sqrt{2 p}$. Using
(\ref{solalphas}), taking into account that $e^{\alpha_+t}$ dies
exponentially like $1/a^{3/2}$, we see that this corresponds to a
Minkowskian mode solution of the standard normalized form
\beq\label{Uplus} u_+(t)=\frac{i}{a\sqrt{k}}e^{-i\frac{k}{a H}}\ . \eeq
When $p$ drops below $\tilde{m}$ we need to match onto the modes in
the $\tilde{m} \gg p$ regime. Notice that in the limit, the mode
$\alpha_{3}$ decays very slowly compared to $\alpha_{\pm}$.  In
essence, these modes have frozen out at the scale $ k = \tilde{m}
a$.

One might have worried that when $V'' > H^2$, the fluctuations of $\phi$ would be massive and would not produce curvature perturbations.  Like in small speed of sound models, we find that the mass can be much larger than $H$ and still contribute to the power spectrum.  Replacing $p^2 \rightarrow p^2+V''$ in (\ref{solalphas}), we find that $\alpha_3 \simeq -3H V''/\tilde{m}^2$.  Therefore, as long as $V'' \ll \tilde{m}^2$, there is still a nearly constant mode that will be converted to curvature perturbations.  This condition is equivalent to (\ref{eta}) and is always satisfied in these models.

When matching the modes at $p \sim \tilde{m}$, it is clear that the
leading terms in $\alpha_{\pm}$ are smooth at the cross-over.  The
real part of $\alpha_\pm$, however, transitions from $-3 H/2$ to $-3H$ in
the crossover between regions I and II. This behavior is distinct
from what would arise for a free scalar field in de Sitter space, and the
matching between the two solutions will introduce new effects
suppressed at small ${\cal O}(H/\tilde m)$.  In order to determine the relative amplitudes of the modes, we cannot simply match the two regimes using continuity at $p \sim \tilde{m}$.  Such a matching calculation assumes that crossover is rapid, but the wavelength $\tilde m^{-1}$ of the modes at the crossover is much smaller than the time period $H^{-1}$ during which the crossover takes place.  Therefore, in order to calculate this sub-leading contributions we will need more than the WKB mode solutions.

Let us therefore move on to discuss the exact solution to the homogeneous linearized equation for the perturbations.
It proves to be convenient to transform the equation to conformal time $\tau=-1/aH$, with late times corresponding to $\tau\to 0$.  Denoting the derivative with respect to $\tau$ by $'$, we have
\beq\label{confintdiff} \varphi''-{2\over\tau}\varphi'+k^2\varphi
+{\hat m^2\over\tau^2H^2}\varphi+{\hat m^2\tau\over H^2}\int_{-\infty}^{\tau}{d\tau'\over\tau'^4}\left({5\over 2}\tau'\varphi'+3\varphi\right)=-\frac{g \Delta n(k,\tau)}{\tau^2 H^2}\ . \eeq
%

Let us comment on the structure of the source on the right hand side of equation~(\ref{confintdiff}).  Since the particle creation happens on very short time scales, we can concentrate on the Minkowski limit. In this case, the squeezed state describing the created $\chi$ particles in the case of homogeneous $\phi$ motion takes the form
\beq\label{squeezed} |\Psi\rangle = {\cal N}\exp\left[\sum_{k_p}{{\beta(k_p) a^\dagger_{\vec k_p} {a^\dagger_{-\vec k_p}}\over{2\alpha^*(k_p)}}}\right]|0\rangle\ ,\eeq
where $\alpha,\beta$ are Bogoliubov coefficients satisfying $|\alpha|^2-|\beta|^2=1$ and ${\cal N}$ is a normalization factor. Here $\vec k_p$ represent the physical momenta, given by $\vec k_p=\vec k/a(t)$, where $\vec k$ is the standard comoving wavenumber. From
this, one computes
the expectation value of the
number density $\int d^3\vec k_p|\beta_{\vec k_p}\beta_{-\vec k_p}|/(2\pi)^3$ given in (\ref{nchi}), using
the standard result (reviewed in \cite{Kofman:2004yc}) that
\beq\label{Bogbeta}\langle a_{\vec k_p}\, a_{\vec k_p}^\dagger \rangle = |\beta(k_p)|^2\sim \exp[-\pi k_p^2/(g\dot\phi)]\ . \eeq
Similarly to the case of the computation of the expectation value of $n_i$, where $i$ represents the $\chi_i$ particle species, it is quite straightforward to see that
\bea
&&\langle\Delta n_i(k,t)\Delta n_j(k',t')\rangle\sim\\ \nonumber&&(2\pi)^3\delta^{(3)}(k+k')\delta_{ij}\frac{(g\dot\phi)^{3/2}}{a(t)^{3/2}a(t')^{3/2}} \Theta(t-t_i)\frac{a(t_i)^3}{a(t)^3}\Theta(t'-t_j)\frac{a(t_j)^3}{a(t')^3}\ .
\eea
Here the $\Theta(t-t_i)$  function  (and analogously $\Theta(t'-t_j)$) represents the fact that, for the population $i$, particle production is irrelevant before the particles become massless.  This is only an approximate expression, which is parametrically correct but that we expect will receive order one corrections in a full calculation. The purpose of this first paper on this class of models is to understand the main features of the predictions, and therefore we consider this level of accuracy enough for the present.
By using the definition
\beq
n(k,t)=\sum_i n_i(k,t)\ ,
\eeq
we obtain:
\bea
&&\langle\Delta n(k,t)\Delta n(k',t')\rangle\sim\sum_{ij}\langle\Delta n_i(k,t)\Delta n_j(k',t')\rangle\sim\\ \nonumber
&&(2\pi)^3\delta^{(3)}(k+k')\sum_i \frac{(g\dot\phi)^{3/2}}{a(t)^{3/2}a(t')^{3/2}} \Theta(t-t_i)\frac{a(t_i)^3}{a(t)^3}\Theta(t'-t_i)\frac{a(t_i)^3}{a(t')^3}\ .
\eea
We can substitute as usual
\beq
\sum_i\simeq\int dt_i \frac{\dot\phi}{\Delta}\ ,
\eeq
to find:
\bea
&&\langle\Delta n(k,t)\Delta n(k',t')\rangle\sim\\ \nonumber
&&\sim(2\pi)^3\delta^{(3)}(k+k')\int dt_i \frac{\dot\phi}{\Delta} \frac{(g\dot\phi)^{3/2}}{a(t)^{3/2}a(t')^{3/2}} \Theta(t-t_i)\frac{a(t_i)^3}{a(t)^3} \Theta(t'-t_i)\frac{a(t_i)^3}{a(t')^3}=\\ \nonumber
&&=(2\pi)^3\delta^{(3)}(k+k')\int^{\min(t,t')} dt_i \frac{\dot\phi}{\Delta} \frac{(g\dot\phi)^{3/2}}{a(t)^{3/2}a(t')^{3/2}} \frac{a(t_i)^6}{a(t)^3 a(t')^3}\ .
\eea
It is straightforward to see that the integral gives:
\bea \nonumber
&&\langle\Delta n(k,t)\Delta n(k',t')\rangle\sim(2\pi)^3\delta^{(3)}(k+k')\frac{(g\dot\phi)^{3/2}}{a(t)^{3/2}a(t')^{3/2}}\frac{\dot\phi}{\Delta H} \frac{a(t_{early})^3}{a(t_{late})^3}\\
&&\label{corrNoDelta}=(2\pi)^3\delta^{(3)}(k+k')\frac{(g\dot\phi)^{3/2}}{a(t)^{3/2}a(t')^{3/2}} N_{hits} \frac{a(t_{early})^3}{a(t_{late})^3}\ .
\eea
where $t_{early}, t_{late}$ are the smaller and greater of $t,t'$.
Here $N_{hits}\sim\dot\phi/(H\Delta)$ is the number of particle production events contributing; because of Hubble dilution, this is limited to events taking place within a Hubble time.


Later in the section, we will see that the $\Delta n$ fluctuations source the inflaton perturbation through the integral in cosmic time of a Green's function whose width in time is of order $H^{-1}$. This means that the inflaton perturbations will be sensitive only to the integral in time of the correlation function of $\Delta n$, and therefore we can approximate the time dependence of the above equation with a $\delta-$function to obtain:
\bea
&&\langle\Delta n(k,t)\Delta n(k',t')\rangle\sim\frac{(g\dot\phi)^{3/2}}{a^{3}(t)}(2\pi)^3\delta^{(3)}(k+k') N_{hits} H^{-1}\delta(t-t')
\eea
We stress that this expression would receive order one corrections in a more exact calculation, but we expect it to capture the correct parametric dependence of the result.

Finally we note that this expression can be obtained more directly in the case where there is a single production event per Hubble time (and correspondingly $N_{hits}$ species in this time).  Then, the particles from the $j$th event have diluted significantly before the next occurs, and the time dependence of the correlation function can be modeled approximately using $j=t_jH$ by $N_{hits}\delta_{ij}=N_{hits}\delta(H(t-t'))=N_{hits}H^{-1}\delta(t-t')$.



It is convenient to rewrite (\ref{confintdiff}) in differential form by acting on it with $\tau{d\over{d\tau}}{1\over\tau}$, giving
\beq\label{confdiffform} \varphi'''+{4\over\tau^2}\varphi'-{3\over\tau}\varphi''+k^2\varphi'
+{\tilde m^2\over{ H^2\tau^2}}\varphi'-{k^2\over\tau}\varphi=-g(g\dot\phi)^{\frac{3}{4}} \sqrt{N_{hits}H^{-1}} \tau{d\over{d\tau}}{\Delta\hat n\over\tau} , \eeq
where $\Delta \hat n$ is defined to have unit variance:
\beq
\langle \Delta\hat n(\eta,k)\Delta\hat n(\eta',k') \rangle=(2\pi)^3\delta^{(3)}(k+k') \delta(\eta-\eta') .
\eeq

In this form, the general homogeneous mode solutions $\varphi_{hom}$ can be written in terms of hypergeometric functions, expandable in terms of Bessel functions.  We find (using Mathematica):
\begin{eqnarray} \label{modesolutions}
\varphi_{hom} &=& c_1\times \,_1F_2\left(-\frac{1}{2};-\frac{i   \tilde m }{2 H}-\frac{1}{2},\frac{i \tilde m }{2 H}-\frac{1}{2};-\frac{1}{4} k^2   \tau^2\right)\nonumber \\
 &+& c_2\times
 2^{-\frac{3 H-i  \tilde m }{ H}} k^{\frac{3 H-i  \tilde m }{ H}}  \, _1F_2\left(1-\frac{i
   \tilde m }{2 H};\frac{5}{2}-\frac{i \tilde m }{2 H},1-\frac{i \tilde m }{H};-\frac{1}{4} k^2
   \tau^2\right) \tau^{\frac{3 H-i \tilde m }{ H}}\nonumber \\
   &+&c_3\times 2^{-\frac{3 H+i \tilde m }{ H}} k^{\frac{3 H+i \tilde m }{ H}}  \,
   _1F_2\left(\frac{i \tilde m }{2 H}+1;\frac{i \tilde m }{2 H}+\frac{5}{2},\frac{i
   \tilde m }{H}+1;-\frac{1}{4} k^2 \tau^2\right) \tau^{\frac{3 H+i \tilde m }{ H}}\nonumber \\
   &\equiv&\sum_{i=1}^3c_if_i(\tau)\ . \end{eqnarray}
The function $f_1$ goes to $1$ as $\tau \to 0$, and it represents the late time constant mode; the other solutions $f_{2,3}(\tau)$ decrease to zero as $\tau\to 0$.
Imposing that this match the Bunch-Davies vacuum solution at early times yields three conditions on the three constants $c_1,c_2$ and $c_3$. We find that for large $\tilde m/H$
\beq\label{Cone}c_1 ~~ \propto  ~~ e^{-\frac{ \tilde m \pi }{2 H}}\ .  \eeq
This leads to a tiny contribution to the power spectrum from homogeneous modes:
\beq\label{Powerhom}P_\zeta^{(hom)}\sim \frac{\left(H^2+ \tilde m^2\right)^2 \text{sech}^2\left(\frac{ \tilde m \pi }{2 H}\right)}{2 {\dot\phi}^2}\times\frac{1}{k^3}\ . \eeq
%
Because of this exponential suppression, the homogeneous contribution will prove to be highly subdominant to the sourced contribution.

To calculate the perturbations generated by the source (\ref{variance}), we must determine the Green's function for the differential equation (\ref{confdiffform}).  We can define $G_k(\tau,\tau')$ as the solution to
\beq\label{confdiffG} G_k'''+{4\over\tau^2}G_k'-{3\over\tau}G_k''+k^2G_k'
+{\tilde m^2\over{ H^2\tau^2}}G_k'-{k^2\over\tau}G_k= \delta(\tau-\tau')\ , \eeq
with the boundary conditions across $\tilde{\tau}=\tilde{\tau'}$ given by
\beq\label{Gbdrycond} \Delta G_k''=1\ , ~~~~ \Delta G_k =\Delta G_k'=0\ .\eeq
It is useful to change variable from $\tau$ to $\tilde\tau=k\tau$, and solve the simpler equation

\beq\label{confdifftildeG}  \dddot{\tilde G}+{4\over\tilde \tau^2}\dot{\tilde G}-{3\over\tilde\tau}\ddot{\tilde G}+\dot{\tilde G}
+{\tilde m^2\over{ H^2\tilde \tau^2}}\dot{\tilde G}-{1\over\tilde\tau}\tilde G= \delta(\tilde \tau-\tilde \tau')\ , \eeq
with the boundary conditions across $\tau=\tau'$ given by
\beq\label{tildeGbdrycond} \Delta \ddot{\tilde G}=1\ , ~~~~ \Delta \tilde G =\Delta \dot{\tilde G}=0\ .\eeq
Here a dot stands for a derivative with respect to $\tilde\tau$. Notice that in this way all the dependence on $k$ is implicit in the definition of $\tilde\tau$, and we have the simple relation:
\beq
G_k(\tau,\tau')=\frac{1}{k^2}\tilde G(k \tau, k\tau')\ .
\eeq
This change of variables will allow us to see analytically that the power spectrum is scale invariant.
The sourced perturbation is given  by
\beq\label{sourcedmodes} \varphi_k(\tau)=g(g\dot\phi)^{\frac{3}{4}} \sqrt{N_{hits}H^{-1}} \int {d\tau'\over\tau'}\Delta\hat n_k(\tau'){d\over{d\tau'}}\left(\tau' G_k(\tau,\tau')\right)\ , \eeq
and the power spectrum at late times ($\tau=0$) is given by
\bea\label{sourcedpowergen}
P_\zeta&=&{H^2\over\dot\phi^2}P_{\varphi}\sim {H^2\over\dot\phi^2}g^{7/2} \dot\phi^{3 /2} N_{hits} H^{-1} \int d\tau'\left[ {1\over\tau'}{d\over{d\tau'}}\left(\tau'G_k(\tau=0,\tau')\right)\right]^2\\ \nonumber
&\sim& \frac{1}{k^3}\times{H^2\over\dot\phi^2}g^{7/2} \dot\phi^{3 /2} N_{hits} H^{-1}\int d\tilde\tau'\left[ {1\over\tilde\tau'}{d\over{d\tilde\tau'}}\left(\tilde \tau'\tilde G(\tilde\tau=0,\tilde\tau')\right)\right]^2\ . \eea
This shows that the power spectrum is scale invariant. In order to determine its amplitude, we need to perform the integral above in (\ref{sourcedpowergen}), where we see that the power spectrum is determined by an `effective' Green's function
\beq\label{gftntilde}\tilde{g}(\tilde\tau,\tilde\tau')\equiv {1\over\tilde\tau'}{d\over{d\tilde\tau'}}\left(\tilde\tau'\tilde G(\tilde\tau,\tilde\tau')\right)=\sum_{i=1}^3f_i(\tilde\tau) \gamma_i(\tilde\tau')\ . \eeq
with only the $f_1(\tilde\tau) \gamma_1(\tilde\tau')$ term surviving as $\tilde\tau\to 0$.
Though we have an analytic expression for the $\gamma_i$'s~\footnote{That we do not reproduce here for the sake of brevity.}, we are unfortunately unable to perform the integral analytically. However, we can notice that the function $\gamma_1(\tilde\tau)$, whose only parametric dependence is on $\tilde m/H$, has a peak at the point $\tilde\tau_*\sim -\tilde m/H$ (corresponding to a physical momentum $p=k/a\sim \tilde m$), with amplitude $\gamma_1(\tilde\tau_*)\sim H^2\tilde\tau_*/\tilde m^2$ and width $\tilde\tau_*$~\footnote{Notice that, as anticipated, in cosmic time, this width corresponds to a time interval of order $H^{-1}$.}.  This allows us to estimate the integral (\ref{sourcedpowergen})  and to obtain the power spectrum:
\beq
\label{power} P_{\zeta} \sim \frac{g^{7/2}H \dot\phi^{1/2}}{\Delta \tilde m}\times \frac{1}{k^3}
\cong  10^{-9}\frac{1}{k^3}\ .
\eeq
This expression can be verified numerically.
\begin{figure}[htp]
\centering
\includegraphics{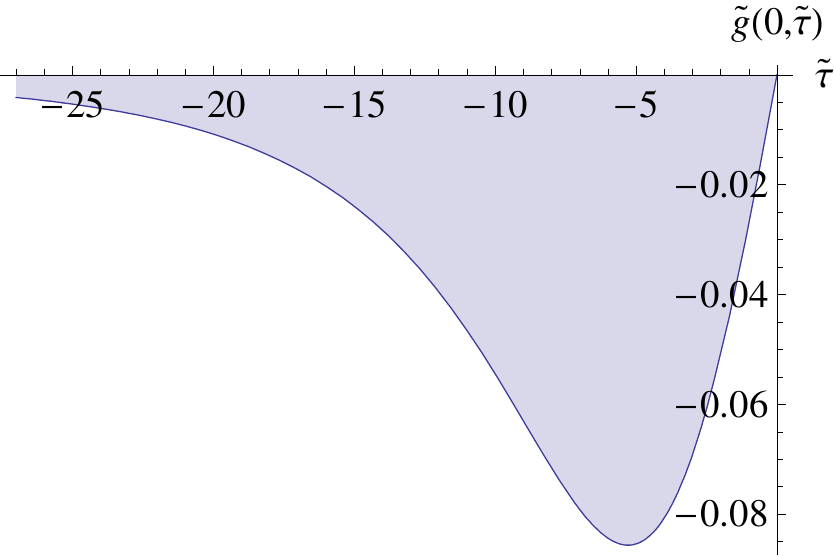}
\caption{A look at the contributions of the Green's function in eq. (\ref{gftntilde}) to the late-time power spectrum, for $\frac{\tilde{m}}{H} = 10$.}
\label{fig:gfn}
\end{figure}
%
%
%
%

Finally, we should ensure that our integral approximation was valid
in this context.  When $p$ is large, it is clear that the variation
of $\varphi$ is large compared to the spacing between particle
production events.  However, the contribution from the integral only
becomes important when the frequency of the modes is $\tilde{m}$.
Therefore, the integral is a good approximation when $ \tilde{m}
\Delta \dot{\phi}^{-1} \ll 1$.  This condition becomes, using our
background solution (\ref{sol}),
\beq \label{pineq} \frac{g^{\frac{3}{2}}
\Delta^{\frac{2}{5}}}{(2\pi)^{9/5}H^{\frac{1}{10}}
V'^{\frac{1}{10}}} \ll 1\ . \eeq

\noindent This is a stronger version of the constraints sketched after eq. (\ref{eom2}).

We derived the integral term assuming that the particle production at each point is the same as for a homogeneous $\phi$ field, which is a valid assumption when the modes of interest obey $p^2 \ll g \dot{\phi}$.  Since the integral term becomes important at the freeze-out scale $p^2 \sim \tilde m^2$, we must have $g \dot{\phi} \gg \tilde m^2$, which gives
\beq \label{deltaconstraint} \frac{(2\pi)^{12/5}\Delta^{4/5}} {g H^{1/5}V'^{1/5}} \gg 1\ . \eeq
This constraint is similar to (and stronger than) eq. (\ref{instant}), but the origins of the two constraints are different.  In the next section, we will look at how all these constraints fit together in a
model with $V = \frac{1}{2}m^2\phi^2$.

Before moving on, let us comment on the role of the $\chi$ fields in the perturbation spectrum.
Our model is not a single field model, given that we require many $\chi$ fields in order
to slow the inflaton.  As such, one might wonder if these extra fields may contribute to
the density fluctuations.  This is not the case because their mass grows to be large before
any modes could freeze out.  Specifically, the effective mass of a $\chi$ field is given
by $m_{eff}^2 = g^2 |\phi(t) - \phi_{i}|^2$.  A Hubble time after the field becomes
massless, the effective mass is given by $m_{eff}^2 \simeq g^2 \dot{\phi}^2 H^{-2}$.  One
can check that $m_{eff}^2 \gg H^2$ is equivalent to our constraint (\ref{instant}).  As
a result, the $\chi$ fields are massive compared to the Hubble scale and do not contribute to the
curvature perturbation~\footnote{It is interesting to consider the fate of these heavy particles.  In some regions of our parameter space, they are always lighter than $M_P$:  $g(\phi_{start}-\phi_{end})\ll M_P$ where $\phi_{start}$ and $\phi_{end}$ refer to the start and end of inflation.  If in other regions they become heavy, they may decay (certainly Planck mass black holes decay rapidly to lighter species of particles).}.

However, given the crucial role of the $\chi$'s in both the background solution and the generation of perturbations, one must ensure that interactions do not cause them to decay.  By construction, the two-body decay $\chi \chi \rightarrow \phi \phi$ is always present.  We can ensure that none of our results are affected by this process by requiring that dilution of particles due to expansion is the primary cause of decreasing number density.  This is expressed by the constraint $n \langle \sigma v \rangle \ll H$.  Assuming $\sigma \propto g^4 (\dot\phi (t-t_i))^{-2}$ and $v \sim {g \dot \phi}^{-1/2} (t-t_i)^{-1}$ we get the condition $ (t - t_i)^3 \gg g^4 (g\dot \phi H)^{-1}$.  Evaluating this expression at the moment the $\chi$ fields are created, $(t_c-t_i )^{-1}= \sqrt{g \dot \phi}$ leads to the constraint
\beq \label{production}
H \gg g^4 \sqrt{g \dot\phi}
\eeq
We will impose this constraint on our parameters although it is possible our results would not be significantly affected even in regions where it is violated.  The mechanism itself can tolerate some $\phi$ production as long as the energy density from the decaying $\chi$'s does not interfere with the perturbations.  

Let us also compare our result for the scalar power (\ref{power}) with the curvature perturbation one obtains from the fluctuations in $\chi$ energy density coming from the variance in $\chi$ particle number on the right hand side of Einstein's equation.  We can estimate this contribution as
\bea \label{potential}
M_{\rm P}^2 \frac{\partial_i^2}{a^2}\zeta_{gravity}\sim m_\chi \Delta n \ .
\eea
Here $\zeta_{gravity}$ is not the curvature perturbation $\zeta$ but comes from the $g_{0\mu}$ components of the metric.  The expression (\ref{potential}) arises from the Hamiltonian constraint.  This contribution is not directly contributing to a measurable power spectrum, but we would like to ensure that the curvature it induces during inflation is negligible.

By going into Fourier space, and using the fact that the
fluctuations are evaluated when $k/a\sim H$~\footnote{This is due to the fact that the fluctuations $\Delta n$ average quickly to zero on scales longer than $H^{-1}$, and therefore the induced metric perturbations become constant after having redshifted up to the scale $H$.}, we obtain, after using
eq. (\ref{corrNoDelta}):
\bea
\langle\zeta_{gravity}(k)\zeta_{gravity}(k')\rangle\sim \frac{1}{k^3}\times
g^{7/2} \frac{\dot\phi^{7/2}}{H^3 M_{\rm P}^4 } N_{hits}\delta^{(3)}(k+k')\
.
\eea
Notice that $m_\chi\sim g \dot\phi H^{-1}$ in this estimate. By comparing with
the contribution we have just computed, $\zeta_\varphi\simeq
\frac{H}{\dot\phi}\varphi$, we obtain
\bea
\frac{\zeta_{gravity}}{\zeta_\varphi}\sim \frac{\dot\phi^2}{V}\left(\frac{\tilde m}{H}\right)^{1/2}\ .
\eea
%
%
%
%
This ratio $\dot\phi^2/V$ is the usual slow roll parameter, which is much smaller than $\epsilon$ in this model. The ratio $\tilde m/H$ has to be smaller than $\sim 10$ because of the constraint coming from non-Gaussianities (see next section).  For the specific case we will study in the next section, where $V(\phi)=m^2\phi^2/2$, the above expression is also equivalent to eq.~(\ref{eta2}) times an additional suppression from the number of e-foldings, and therefore it is always satisfied in that model.

\subsection{Non-Gaussian Perturbations}

The size and shape of the non-Gaussian contribution to the
perturbations are particularly important for distinguishing between
different models of inflation \cite{Babich:2004gb}.  Since our interactions slow the inflaton on a potential which would otherwise be too steep for inflation, we should expect a substantial non-Gaussian correction to the power spectrum as in \cite{Alishahiha:2004eh}.  A detailed prediction for the bispectrum requires the calculation of the three-point correlation function of the curvature perturbation, as first completed for single-field slow roll inflation in \cite{Maldacena:2002vr,Acquaviva:2002ud}.


Following \cite{warmperts}, we can expand the equation of motion (\ref{expandedeom}) for the $\varphi$ perturbation into first order, second order, and higher order pieces:
\beq\label{orders} \varphi\equiv\varphi^{(1)}+\varphi^{(2)}+\dots\ . \eeq
It is again useful to translate the expanded equation of motion into conformal time, and derive its differential form (as done for the linearized equation in (\ref{confdiffform})).  Then we can obtain the second order perturbation $\varphi^{(2)}$ by integrating against the Green's function $G_k(\tau,\tau')$ the terms in the expanded equation of motion which are second order in $\varphi^{(1)}$ and $\Delta n$.  By looking at eq. (\ref{expandedeom}), one sees that one contribution comes from the expansion of the term proportional to $\tilde m^2$ in the equation of motion, giving a contribution to the second order perturbation of order:
\beq\label{NGI} \varphi^{(2)}_{k,\tilde m}(\tau)\sim \int d\tau' G_k(\tau,\tau'){\tilde m^2\over{H\tau'\dot\phi}}
(\varphi'(\tau')\varphi'(\tau'))_k\ . \eeq
Another contribution comes from taking into account the time delay of the perturbation inside the integral, giving rise to a term of the form:
\beq\label{NGIII} \varphi^{(2)}_{k,\delta t}(\tau)\sim \int d\tau' G_k(\tau,\tau'){\tilde m^2\over{H\tau'\dot\phi}}
(\varphi''(\tau')\varphi(\tau'))_k\ . \eeq
Yet another contribution, of order $\varphi^{(1)}\Delta n$, comes from expanding the $(g\dot\phi)^{3/4}$ coefficient in the source term, giving a contribution
\beq\label{NGII} \varphi^{(2)}_{k,\Delta n}(\tau)\sim g^{7/4} \dot\phi^{-\frac{1}{4}} \sqrt{N_{hits}H^{-1}} \int d\tau'G_k(\tau,\tau')\tau'{d\over{d\tau'}}{(\dot\varphi^{(1)}(\tau') \Delta \hat n(\tau'))_k \over\tau'} \ .\eeq
There are additional terms coming from the expansion of $\tilde{m}^2$, but it is easy to see that they give subleading contributions. Also the contribution from the non-Gaussian statistics of $\Delta n,$ which in the absence of interactions can still come from $\chi$-particle shot noise, is expected to be negligible if the number of particles is large enough.  This is in fact always the case. Estimating the size of the non-gaussianity of $\Delta n$ by $(\,\tilde{m}^3/(n_{\chi}N_{hits})\,)^{1/2}$, where we have used that $\tilde m$ is the typical scale at which the Green's functions peak, it is easy to see that in our model, by using eq.~(\ref{power}), this ratio is smaller than $10^{-6}$, corresponding approximately to a negligibly small $f_{\rm NL}\sim 0.1\; $.


The three point function of our perturbations is of the form
\beq\label{threept}  \langle \varphi^{(2)}_{k_1}(\tau)\varphi^{(1)}_{k_2}(\tau)\varphi^{(1)}_{k_3}(\tau)\rangle \ ,\eeq
and we are interested in this amplitude at late times, $\tau\to 0$.
We can estimate this using the same method we used for the Gaussian power spectrum, and let us start with the term in eq.~(\ref{NGI}).   The perturbations on the right hand side of (\ref{NGI}) can come from any of the three modes (\ref{modesolutions}), not only from the constant mode $f_1$
.  This is so because the perturbations in $\varphi^{(1)}$ that source the second order $\varphi^{(2)}$ in eq.~(\ref{NGI}) can be evaluated when still well inside the horizon when neither of the three modes has yet decaied. One of the leading effects we find comes from the $f_2$ and $f_3$ modes~\footnote{A similar term with a pair of $f_2$ or a pair of $f_3$ modes will change the final result by no more than an $O(1)$ factor.  We study the $f_2 f_3$ term above because the cancellation between phases is particularly simple.}, giving a contribution to the three point function of curvature perturbations of order
\bea\label{threept23}
&&(2\pi)^3\delta^{(3)}(\sum_{i=1}^3\vec k_i)\left({H\over\dot\phi}\right)^3\int d\tau\,G_{k_1}(0,\tau) {\tilde m^2\over{\tau H\dot\phi}}\left(g(g\dot\phi)^{\frac{3}{4}} \sqrt{N_{hits}H^{-1}}\right)^4 \\
&&\times \int^\tau d\tau' g_{k_2}(0,\tau')g_{2,k_2}'(\tau,\tau')\int^\tau d\tau'' g_{k_3}(0,\tau'')g_{3,k_3}'(\tau,\tau'')+{\rm symm.}\  .
\eea

If we pass to the Green's functions $\tilde{G},\tilde{g}$ defined as in the former section, we find:
\bea\label{threept23tilde}
&&
 (2\pi)^3\delta^{(3)}(\sum_{i=1}^3\vec k_i)\left({H\over\dot\phi}\right)^3\int \frac{d\tilde\tau}{k_1}\,\frac{\tilde G(0,\tilde\tau)}{k_1^2} {k_1 \tilde m^2\over{\tilde\tau H\dot\phi}}\left(g(g\dot\phi)^{\frac{3}{4}} \sqrt{N_{hits}H^{-1}}\right)^4 \\ \nonumber
&&\times \int^{\tilde\tau k_2/k_1} \frac{d\tilde\tau'}{k_2} \frac{\tilde g(0,\tilde\tau')}{k_2} \frac{\tilde g_{2}'(\tilde \tau k_2/k_1,\tilde\tau')}{k_2}k_2 \int^{\tilde\tau k_3/k_1} \frac{d\tilde\tau''}{k_3} \frac{\tilde g(0,\tilde\tau'')}{k_3} \frac{\tilde g_{3}'(\tilde \tau k_3/k_1,\tilde \tau'')}{k_3}k_3+{\rm symm.}= \\ \nonumber
&&= (2\pi)^3\delta^{(3)}(\sum_{i=1}^3\vec k_i)\left({H\over\dot\phi}\right)^3\left(g(g\dot\phi)^{\frac{3}{4}} \sqrt{N_{hits}H^{-1}}\right)^4\frac{1}{k_1^6}\frac{1}{x_2^2 x_3^2}\int d\tilde\tau\,\tilde G(0,\tilde\tau) { \tilde m^2\over{\tilde\tau H\dot\phi}} \\ \nonumber
&&\times \int^{\tilde\tau x_2} d\tilde\tau' \tilde g(0,\tilde\tau') \tilde g_{2}'(\tilde \tau x_2,\tilde\tau') \int^{\tilde\tau x_3} d\tilde\tau'' \tilde g(0,\tilde\tau'') \tilde g_{3}'(\tilde \tau x_3,\tilde \tau'')+{\rm symm.}\ .
\eea
where we have defined $x_2=k_2/k_1$ and $x_3=k_3/k_1$.
The former expression is of the form
\bea
&& (2\pi)^3\delta^{(3)}(\sum_{i=1}^3\vec k_i)\left({H\over\dot\phi}\right)^3\left(g(g\dot\phi)^{\frac{3}{4}} \sqrt{N_{hits}H^{-1}}\right)^4\frac{1}{k_1^6} {{\cal G}(x_2,x_3,\tilde m/H)}+{\rm symm.}\\ \nonumber
&&\equiv (2\pi)^3\delta^{(3)}(\sum_{i=1}^3\vec k_i)F(k_1,k_2,k_3)
\eea
The factor of $1/k^6$, which characterizes the dependence on the global scale of the momenta, tells us that the signal is scale invariant \cite{Babich:2004gb}.


As we discussed above, the Green's functions $G(0,\tau)$ or $g(0,\tau)$ are peaked at $\tau_*\sim -\tilde m/(Hk)$, and the product of integrals forming the $\tau$ integrand also exhibit a peak at this value.  We can estimate the size of $F(k_1,k_2, k_3)$ using knowledge of the peak at $\tau_*$ and series expansion of the Green's functions around $\tau = 0$.
We find that the Green's functions $g(0, \tilde{\tau})$ and $g_{2,3} (\tilde{\tau},\tilde{ \tau'})$ can be expanded as a series of the form  $\tilde{\tau}^{p} \sum a_n(\tilde{\tau}/\tilde{\tau}_*)^n$ with order one coefficients $a_n$.  Physically, we believe this occurs because the only features of these functions occur near $\tilde{\tau} \sim \tilde{\tau}_*$.  Therefore, we only expect any non-trivial behavior when $\tilde{\tau} \sim \tilde{\tau}_*$. Using the above Taylor expansion at $\tau\sim\tau_*$ is likely inaccurate, but we think it should be reliable for order of magnitude estimates~\footnote{All the results using series expansions have been checked against numerical integrations and provide reliable estimates.}.

\begin{figure}[htp]
\centering
\includegraphics[scale =0.75]{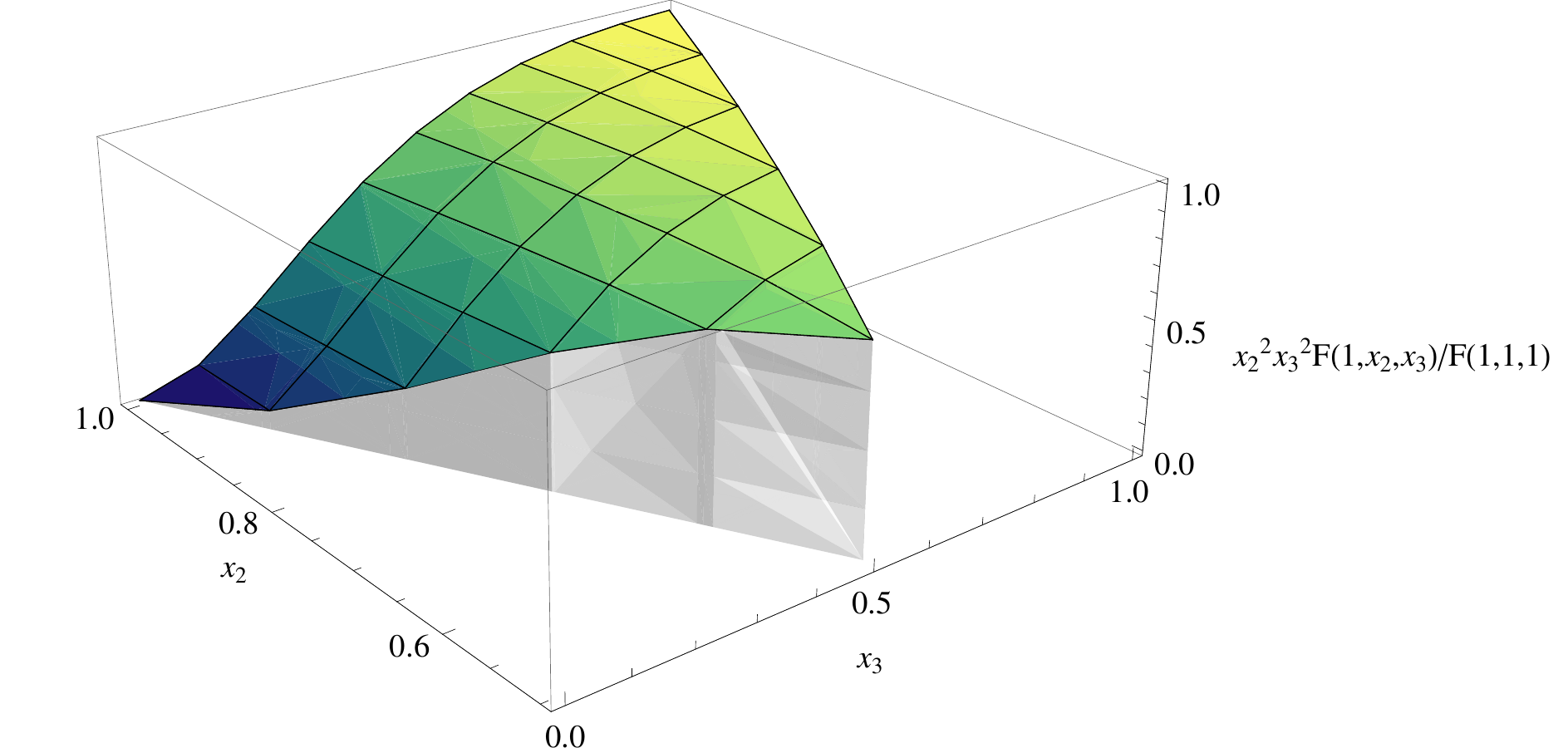}
\caption{A numerical study of the shape $x_2^2 x_3^2 F(1,x_2,x_3)\over F(1,1,1)$ for the choice of parameters $\frac{\tilde{m}}{H} = 10$, plotted in the region $0 \leq x_2 \leq 1; 1 - x_2 \leq x_3 \leq x_2$.  The peak in the equilateral limit is clearly visible.}\label{fig:fplot}
\end{figure}

In the language of eq. (\ref{gftntilde}), the leading terms in the expansion for $f_{2,3}(\tilde{\tau})$ are of order $\tilde{\tau}^3\tilde{\tau}^{\pm i\frac{\tilde{m}}{H}}$ and those for $\gamma_{2,3}(\tilde{\tau})$ are of order $\frac{H}{\tilde{m}}\frac{1}{\tilde{\tau}^2}\tilde{\tau}^{\mp i\frac{\tilde{m}}{H}}$.  The oscillations contribute a suppression factor $\frac{H}{\tilde{m}}$ to the integrals and an $\frac{\tilde{m}}{H}$ enhancement to the derivatives. Altogether this gives us an estimate, which we checked against a numerical integration, of order
\beq\label{threeptvalue} \langle\zeta_{\vec k_1}\zeta_{\vec k_2}\zeta_{\vec k_3}\rangle\sim (2\pi)^3 \delta^{(3)}(\sum_{i=1}^3\vec k_i) {H^2 g^7 N_{hits}^2\over{k^6\dot\phi}}\ . \eeq
Here we were not careful with the momenta dependence, and the factor $k$ denotes only the typical size of the wavenumber.  A more careful numerical analysis for the shape function $x_2^2 x_3^2 F(1,x_2,x_3)/F(1,1,1)$ as defined in \cite{Babich:2004gb} shows that most of the signal is concentrated on equilateral configurations.  The equilateral shape can be understood to be a result of the Green's functions being peaked at a scale $\tau_{*}$: we get a large contribution when all the momenta are equal and all the Green's functions can be evaluated at their peak value.  More in detail, by looking at eq.~(\ref{threept23tilde}), one can notice that in order for the integrals in $\tilde\tau'$ and $\tilde\tau''$ to include in their domain the peaks of $\tilde g(0,\tilde\tau')$ and of $\tilde g(0,\tilde\tau'')$ by the time $\tilde G(0,\tilde\tau)$ reaches its peak at $\tilde\tau\sim\tilde\tau_{*}$, we need to have $x_2, x_3\lesssim 1$. However, in the limit $\tilde\tau \rightarrow 0$, we have approximately $\tilde g'_{2,3}(\tilde \tau x_{2,3},\tilde  \tau') \propto\tilde \tau^2 x_{2,3}^2$, which suppresses this contribution to the shape by $x_2^2x_3^2$ and forces the dominant contribution to come from the case where $x_2, x_3$ are as large as possible compatibly with the former constraint. We obtain that the integrals are peaked for  $x_2,x_3\simeq1$, on equilateral configurations~\footnote{There is some support also on flattened triangles, but  the numerical study plotted in Fig.~\ref{fig:fplot} shows that this does not dominate over the equilateral shape.}. The suppression of $\tilde g'_{2,3}(\tilde \tau,\tilde  \tau')$ at small $\tilde\tau$ comes from the fact that the oscillating modes decay at late time.  Notwithstanding the fact that the leading mechanism for generating non-Gaussianities is intrinsically a multifield effect, we conclude that the signal on squeezed configurations is not large, as is always the case in single field inflation \cite{Maldacena:2002vr,Acquaviva:2002ud,Creminelli:2003iq,Creminelli:2004yq,Cheung:2007sv}.

A similar analysis shows that the contribution due to $\varphi^{(2)}_{k,\delta t}(\tau)$ is parametrically the same as the one of $\varphi^{(2)}_{k,\tilde m}(\tau)$, while the one from $\varphi^{(2)}_{k,\Delta n}(\tau)$ is suppressed by a factor of ${H / \tilde{m}}$. The remaining terms that we did not show are subleading as well. Since we are not careful with order one coefficients, there is no need to perform the calculation for $\varphi^{(2)}_{k,\delta t}(\tau)$, since we do not expect cancellations or the shape to be peaked in the squeezed limit.


Summarizing, following the standard definition, we can estimate the size $f_{\rm NL}$ on equilateral triangles (with $|\vec k_i|\equiv k$) to be of order
\beq\label{fnl} f_{\rm NL}^{\rm equilateral}\sim {\langle\zeta_{\vec k_1}\zeta_{\vec k_2}\zeta_{\vec k_3}\rangle'
\over{\langle\zeta_{\vec k}\zeta_{-\vec k}\rangle^{'2}}}\sim {\tilde m^2\over H^2} \ . \eeq
where the primes indicate that we dropped the delta functions of momenta.

\section{The case $V(\phi)= \frac{1}{2} m^2 \phi^2$}

Let us now check the conditions for a viable model of trapped inflation, including the background solution and Gaussian perturbations.  We will take a model with potential $V(\phi)={1\over 2}m^2\phi^2$ for simplicity; other cases
of interest include more general power law potentials $V_\alpha(\phi)=\mu^{4-\alpha}\phi^\alpha$.
Given the number of e-foldings, the Gaussian power spectrum (\ref{power}), and our solution (\ref{sol}),
we can solve for two of the parameters and then express the various inequalities prescribed in \S2\ in terms of fewer model parameters.  From this, we obtain the following relations.

\begin{figure}[htp]
\centering
\includegraphics[width=11cm]{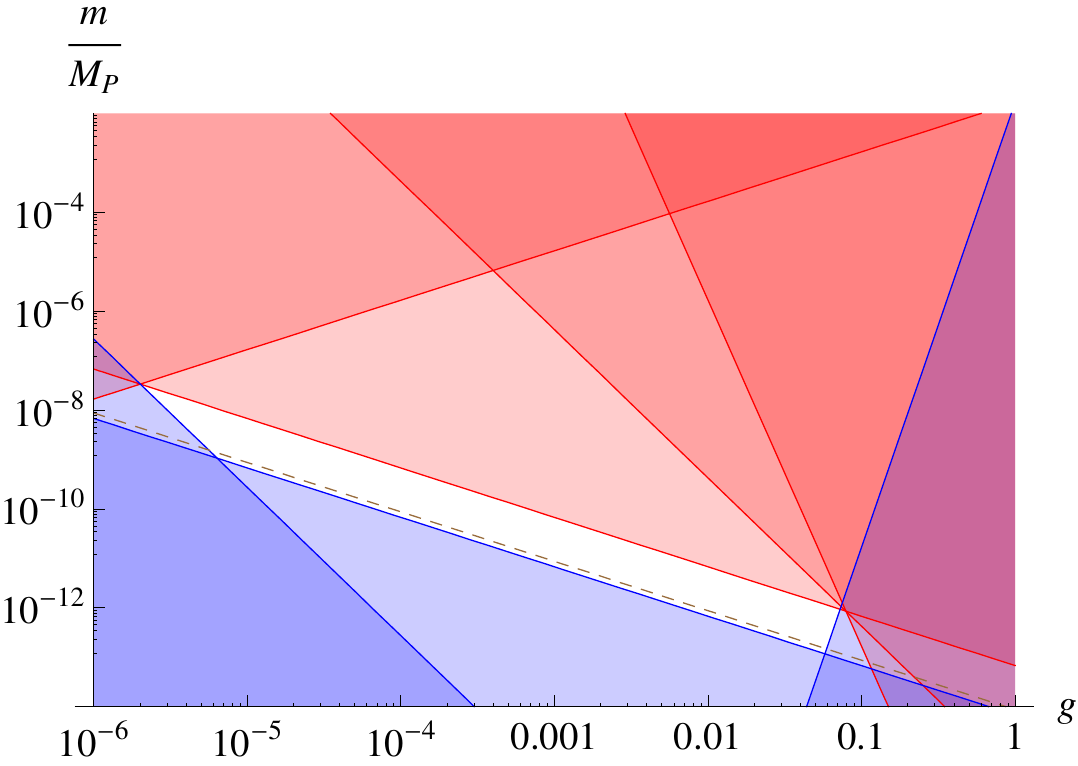}
\includegraphics[width=11cm]{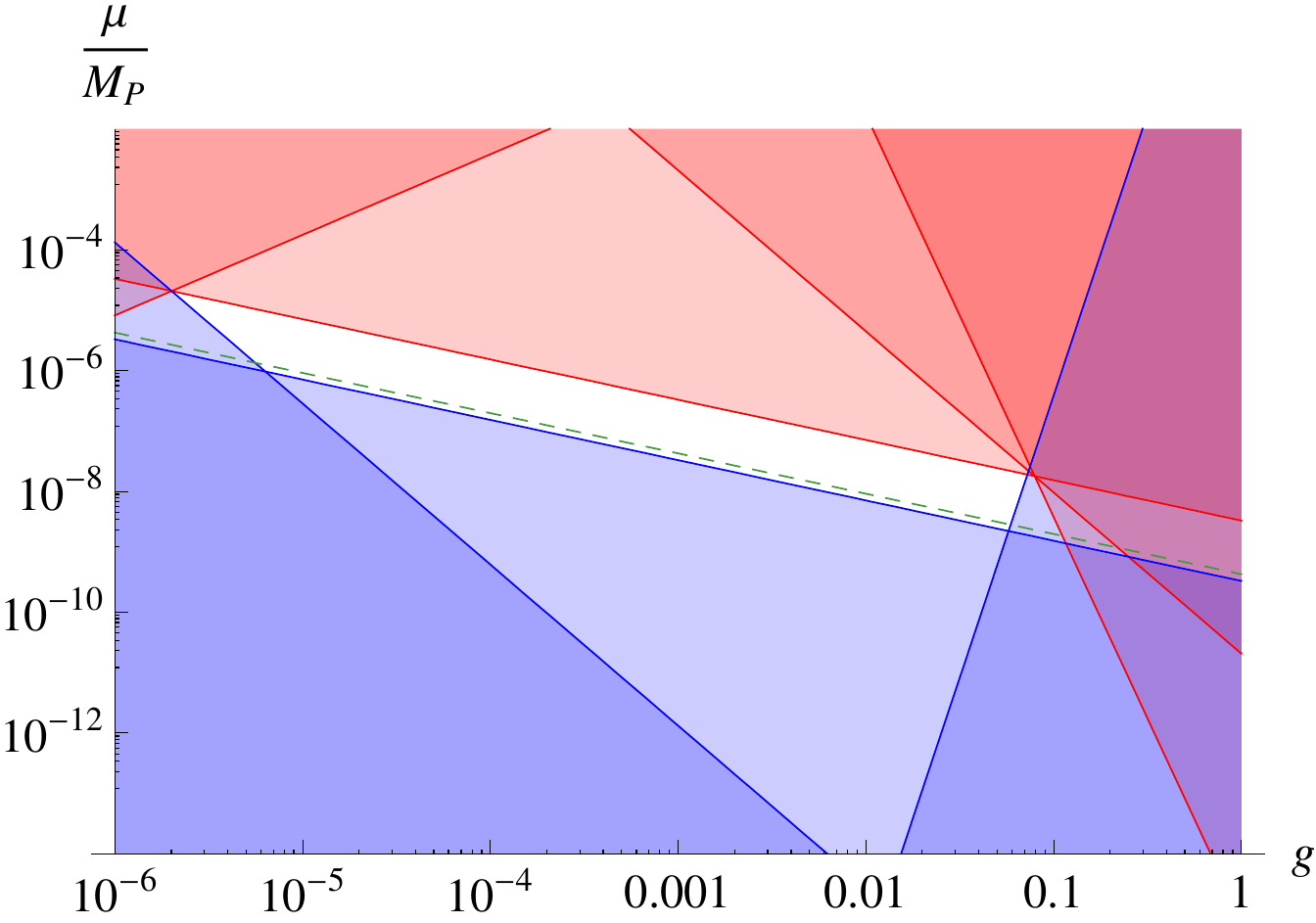}
\caption{{\it Top:} The allowed parameter window for the $m^2 \phi^2$ model.  The red zones are forbidden by eq. (\ref{conditions}); the blue zones are forbidden by eq. (\ref{deltacondition}) and by the constraint (\ref{NGconstraintm}) on the size of the non-Gaussianities, to be discussed in the next section.  The dashed line indicates the range of parameters for which $\phi/M_P\sim 1$, with super-Planckian field ranges above and sub-Planckian ranges below. {\it Bottom:} Same plot as above for a model with potential equal to $\mu^3\phi$. We do not give explicitly the constraints in the paper as they are very similar to the ones for the $m^2\phi^2$ model.}\label{fig:consistwindow}
\end{figure}

The number of e-foldings is
\beq\label{Ne} N_e=\int {H\over\dot\phi}d\phi \sim {10^{-6}\over{(2\pi)^{2}g^{2/3}}}\left({M_P\over m}\right)^{2/3}\left({\phi\over M_P}\right)^{2/3}\ = \frac{10^{-6} }{g^{2/3} (2\pi)^2}  \left({\phi \over m}\right)^{2/3} , \eeq
where the slow roll condition (\ref{ineq1}) forced the total field range $\phi_{i} - \phi_{f} \simeq \phi_{f} \equiv \phi$ and we used (\ref{power}) to eliminate $\Delta$.  Using this, (\ref{power}), and our solution (\ref{sol}) to write the self-consistency conditions in terms of $m/M_P$, the constraints  (\ref{ineq2}), (\ref{ineq3}), (\ref{eta2}), (\ref{pineq}) respectively give four conditions~\footnote{The constraints (\ref{ineq1}), (\ref{eta}) give $\frac{1}{N_e} \ll 1$, and so are trivially satisfied.}
\begin{eqnarray}\label{conditions} {mN_e\over{M_P g}}\ll 1\ , &\qquad&
(2\pi)^{3}10^{18}g^{9/2} N_{e}^{3/2} \left({m\over M_P}\right)^{3/2}\ll 1\ , \nonumber\\
  (2\pi)^{6}10^{18}N_e^2 g^2\left({m\over M_P}\right)^{2} \ll 1 \ ,  &\qquad& \ 10^9 g^{7/2} N_{e}^{1/2} \left({m\over M_P}\right)^{1/2} \ll 1. \end{eqnarray}
The constraint (\ref{deltaconstraint}) and (\ref{production}) are in the other direction:
\beq \label{deltacondition} g^3 10^{18} (2\pi)^6 N_e \left({m\over M_P}\right) \gg 1, \qquad  g^{-9} N_e \left({m\over M_P}\right) \gg 1 \ . \eeq
Together, these constraints define a viable window in the space defined by the two free parameters $(g, {m\over M_P})$, which is plotted in Fig.~\ref{fig:consistwindow}.  Maximizing the value of the field range $\phi\over M_P$ over this window, the field range is constrained to lie no more than an order of magnitude above the Planck scale.  The mechanism therefore can operate below the scale typically needed for standard slow-roll inflation - in fact the constraints described above allow field values far below $M_P,$ although we will see in the next section that experimental constraints on the size of the non-Gaussianity in the power spectrum prevent us from going far below the Planck scale in this model.

\section{Observational predictions}

In this section we will outline the predictions for the CMB derived from our inflationary mechanism.

\subsection{$n_s$ and $r$}

Because $H$ and other background parameters change slowly during inflation, our power spectrum is approximately scale invariant.  Its tilt is given by %
\beq\label{tiltgen} n_s-1 = \frac{d {\rm ln} P_\zeta}{d{\rm ln} k}\approx \frac{d {\rm ln} P_\zeta}{H dt} \ .\eeq
From (\ref{power}) this becomes
\beq\label{tiltII} n_s=1+ \frac{\dot H}{H^2}-\frac{\ddot\phi}{4H\dot\phi}\ \simeq1-\frac{0.7(1-Q^{6/5})}{N_e}\simeq 0.99 \ .
\eeq
where in the next to last passage we used eqs.~(\ref{Ne})~and~(\ref{sourcedpowergen}), and in the last passage we have used a typical number of e-foldings $N_e\sim 55$ \footnote{For potentials of the form $\mu^{4-\alpha}\phi^\alpha$, $n_s-1=-(2+7\alpha)(1-Q^{7/5-\alpha/10})/ (2N_e (14-\alpha))$  .}. The parameter $Q$ we have introduced here represents the ratio $\phi_{\rm end}/\phi_{N_e}$, between the value of $\phi$ at $N_e$ efoldings to the end of inflation and the value at the end of inflation. This parameter was not introduced in the former estimates because it does not affect them significantly.  The condition $\epsilon \ll 1$ requires that $Q^{-6/5}-1 \ll  N_{e}$, so we can safely take $Q \sim 1/4$.  The tilt is red, falling quite close to the statistically preferred region of the WMAP 5-year  data \cite{Observations}.  However, it is worth mentioning that the tilt may not be a sharp prediction of this class of of models, but may be tunable in
general.  It can depend not only on the potential and the field range, but also on other details such as variation in the spacing between the particle production events, and in the mass and
species numbers of of the particles.  In order to compute the tilt very precisely, it would also be important to systematically check the contributions of higher dimension operators.  These are limited by
symmetries in our string-theoretic backgrounds, but we have not done a complete analysis of their leading effects.

The power in gravity waves is as usual $P_{tensor}\approx \frac{8}{M_P^2}\left(\frac{H}{2\pi}\right)^2$, leading to a tensor to scalar ratio of:
\beq\label{r} r=\frac{P_{tensor}}{P_\zeta}\approx \frac{\Delta \tilde m H}{g^{7/2} \dot{\phi}^{1/2} M_P^2} \sim g^2 N_e^3 10^{27} (2\pi)^6 \left({m\over M_P}\right)^4\ . \eeq
in the ${1\over 2}m^2 \phi^2$ model.  Maximizing this quantity over the allowed range of $(g, {m\over M_P})$ from the previous section, we find $r \ll 10^{-4}$ for this potential.
%
%

\subsection{Non-Gaussianity}

Current constraints from data \cite{Observations, Creminelli:2006rz}\ bound $f_{\rm NL}^{\rm equilateral}$ so that, using eq.~(\ref{fnl}), we have
\beq\label{NGconstraint} {\tilde m\over H} \lesssim 10\ . \eeq
Note that since we have not been keeping track of $O(1)$ factors, there is a possibility that these may shift this constraint slightly in either direction.

Plugging this into our solution, this is equivalent to
\beq\label{NGconstraintII} {V'\over{(2\pi)^{3}g H^3}} < 10^{12} \ .\eeq
For $m^2\phi^2$ this corresponds to
\beq\label{NGconstraintmphi} {1\over (2\pi)^3 g}\left({M_P\over\phi}\right)^2\left({M_p\over m}\right) < 10^{12}\ , \eeq
and imposing (\ref{Ne}) we obtain
\beq\label{NGconstraintm} g{m\over M_P}> {10^{-10}\over{N_e (2\pi)^{3}}} \ .\eeq
This goes in the opposite direction from the previous conditions (\ref{conditions}) (except for (\ref{deltacondition})), but leaves a wide window of viability.  As seen in Fig.~\ref{fig:consistwindow}, it is the constraint on the non-Gaussianity that restricts the field range from going far below the Planck scale.  This is to be expected - as in \cite{Silverstein:2003hf},\cite{Alishahiha:2004eh}, as the potential grows steeper a stronger interaction will be needed to slow the inflaton, and a larger contribution to the non-Gaussianity will be produced.

%
%
%
%

\section{Trapped Inflation from String Theory}
\label{sec:string}

\smallskip

\noindent {\it ``Meetings are a great trap..."} ~~  --John Kenneth Galbraith

\bigskip


Because inflation is sensitive to Planck-suppressed operators in the effective field theory, it is generally of interest to model it in a UV complete theory of gravity.  String theory, as a candidate UV completion of gravity, is a standard framework in which to develop such constructions.  The present work was motivated in part by the top-down appearance of the structure required for trapped inflation.  In this section, we will explain this structure and analyze the conditions for realizing trapped inflation consistently with moduli stabilization in appropriate examples.  These realizations use the same structures recently used for monodromy-driven large field inflation \cite{Silverstein:2007ac,McAllister:2008hb}, but now in a $\lesssim M_P$ range of field.  Because the relevant setups were described in detail in these works, our discussion here will be somewhat more telescopic; the reader may therefore find it easiest to refer back to the relevant portions of \cite{Silverstein:2007ac,McAllister:2008hb}.

To begin, consider wrapped D4-branes in type IIA string compactifications on nilmanifolds, as in \cite{Silverstein:2008sg,Silverstein:2007ac}. The simplest example of a Nil manifold suffices to exhibit our basic mechanism for closely spaced particle production events, though we will see that trapped inflation in this specific example would introduce too large a back reaction on the internal geometry.  We will therefore ultimately be led to construct it in string theory by using axion moduli in warped Calabi-Yau compactifications of the kind analyzed recently in \cite{McAllister:2008hb}.  Particle production in these models was also considered in \cite{Brandenberger:2008kn} where it was used for reheating.

A nil 3-manifold is obtained by compactifying the nil geometry
\begin{eqnarray} \label{nilgeom} ds^2_{Nil} &=& \frac{L_{u}^2}{\beta}du_1^2+\beta L_{u}^2du_2^2 + L_x^2\left(dx+\frac{M}{2}[u_1 d u_2-u_2 d
u_1]\right)^2 \nonumber \\
&=& \frac{L_{u}^2}{\beta}du_1^2+\beta L_{u}^2du_2^2 + L_x^2\left(dx'+{M}u_1 d u_2\right)^2\ ,
\end{eqnarray}
(where $x'= x-\frac{M}{2}u_1u_2$) by a discrete subgroup of the isometry group
\begin{eqnarray} \label{translations}
 t_x: ~~ (x,u_1,u_2) &\to & (x+1, u_1,u_2)~, \nonumber \\  t_{u_1}: ~~(x,u_1,u_2) &\to &
 (x-\frac{M}{2}u_2, u_1+1,u_2)~, \nonumber \\ t_{u_2}:
~~ (x,u_1,u_2) &\to & (x+\frac{M}{2}u_1,u_1,u_2+1)\ . \end{eqnarray}
This manifold can be described as follows.  For each $u_1$, there is a torus in the $u_2$ and
$x'\equiv x-\frac{M}{2}u_1u_2$ directions. Moving along the $u_1$ direction, the complex structure $\tau$ of
this torus goes from $\tau\to \tau +M$ as $u_1\to u_1+1$. The projection by $t_{u_1}$ identifies these
equivalent tori~\footnote{The directions $u_1$ and $u_2$ are on the same footing; similar statements apply with
the two interchanged and with $x'$ replaced by $x''\equiv x+\frac{M}{2}u_1u_2$.}.

At all values $u_1=j/M$ for integer $j$, the two-torus in the $u_2-x'$ directions is equivalent to a rectangular
torus
\be ds^2_{rect}\equiv L_x^2 dy_1^2+\beta L_{u}^2dy_2^2\ , ~~~~~ (y_1,y_2)\equiv (y_1+n_1,y_2+n_2)\ , \ee
(since $\tau\to\tau+1$ as $j\to j+1$).  These coordinates $y_1$ and $y_2$ are related to $x'$ and $u_2$ by an
$SL(2,Z)$ transformation. The 1-cycle traced out by $u_2=\lambda,\lambda\in (0,1)$  becomes a cycle
$(y_1,y_2)=(M\lambda,\lambda)$ as $u_1\to u_1+1$.

Consider first, as in \cite{Silverstein:2008sg}, a D4-brane wrapped on this cycle.  Near $u_1=0$, it has a potential energy of the form
\beq\label{nilpot} V(\phi)=\frac{1}{2}m^2\phi^2\ , ~~~~~~~ \phi \le M_P \ ,\eeq
in terms of the canonically normalized field $\phi$ corresponding to its collective coordinate in the $u_1$ direction.  This collective coordinate will play the role of the inflaton, and we will refer to this D4-brane as the inflaton brane.

As mentioned above, at $u_1=j/M, j=1,\dots, M$ there is a rectangular torus in the $u_2,x'$ directions, equivalent by an SL(2,Z) transformation to the one at the origin.
Introduce $N_4$ additional D4-branes wrapped on the corresponding SL(2,Z) transforms of the cycle wrapped by the inflaton brane.  The $jth$ such brane has a quadratic potential proportional to $(u_1-j/M)^2$, minimized at $u_1=j/M$.  Place each at its minimum.  As the inflaton brane rolls down its potential (\ref{nilpot}), it encounters these additional branes, causing the strings $\chi_j$ (and fermion partners) stretched between them to come down to zero mass.  That is, $\phi$ and the $\chi_j$ couple as in our basic field theory model (\ref{interaction}).

It is clear that this structure arises more generally than the particular model \cite{Silverstein:2007ac,Silverstein:2008sg}.  In this particular case it is worthwhile to analyze the consistency of these added branes with the moduli stabilization barriers introduced by the curvature of the nilmanifold and other ingredients required to stabilize the space.  A single D4-brane at the minimum of its potential is subdominant to the moduli-stabilizing barriers.  There is a limit to how many additional branes can coexist with moduli stabilization.  The tension of the set of D4-branes is
\beq\label{totalBpot} V_{D4}= N_4 \frac{\sqrt{\beta}L_u}{(2\pi)^4 g_s\alpha'^2}\ . \eeq
This must be less than the scale of the moduli-stabilizing barriers, of order the curvature-induced potential energy:
\beq\label{VlessU} V_{D4}<\frac{L_x^4M^2}{(2\pi)^7g_s^2\alpha'^2}\ . \eeq
Now in terms of the field theoretic quantities of the previous sections, $N_4\sim \phi/\Delta$.  So the condition (\ref{VlessU}) translates into the condition
\beq\label{Nfourbound} N_4\sim \frac{\phi}{\Delta} < \frac{L_x^4M^2}{(2\pi)^3g_s L_u\sqrt{\beta}} \ .\eeq
In the simplest version of the construction \cite{Silverstein:2007ac}\ -- with the numerical examples discussed there and in \cite{Silverstein:2008sg}\ -- the number of D4-branes is limited by this back reaction to be of order 10.
Possibilities for warping down excessive contributions to the potential energy were discussed in \cite{Silverstein:2007ac}.  In general, the mechanism we have discussed arises in a wide variety of ``monodrofold" type compactifications \cite{monodrofold}.

A similar structure, with somewhat more flexibility in the parameters, arises in the setting \cite{McAllister:2008hb}\ to construct trapped inflation from string theory.  Consider type IIB string theory on a warped Calabi-Yau manifold, with an axion $c$ arising from a 2-form RR potential $C^{(2)}$ integrated over a 2-cycle $\Sigma_2$.  In the presence of an NS5-brane wrapped on $\Sigma_2$ within a warped region (with a corresponding anti-brane wrapped on a homologous cycle in a distant warped region), the potential for $c$ takes the form
\beq\label{cpot} V(c)= {\epsilon\over{g_s^2(2\pi)^5\alpha'^{2}}}\sqrt{\ell^4+c^2
g_s^2} \ ,\eeq
where $\epsilon$ encodes the warp-factor dependence.  As explained in \cite{McAllister:2008hb}, the axion decay constant is of order $f\sim M_P/L^2$.   This setup, with a large stabilized 2-cycle size $\ell$, naturally realizes large-field inflation (with $c$ playing the role of the inflaton, executing many cycles of its basic period $c\to c+(2\pi)^2$).  For the case of a blown-down 2-cycle, $\ell\to 0$, the same setup leads to trapping as follows.  When $\ell=0$, as $c$ rolls through the values $(2\pi)^2 j$ (with $j$ an integer), new light degrees of freedom appear in the theory.  One intuitive way to see this is via the S- and T- dual setup depicted in Fig. 1 of \cite{McAllister:2008hb}\ -- there the NS5-branes' horizontal separation corresponds to $\ell$, and when this vanishes the unwinding motion takes the system through configurations where these NS5-branes meet.  At these points, new light degrees of freedom arise from stretched D2-branes; the theory at low energies is a nontrivial interacting CFT (see e.g. \cite{Seiberg:1997td}).  The massless degrees of freedom of this CFT are produced much in the same way as are the $\chi$'s described above (though perhaps in this case we should call it {\it un}particle production, since the low-lying degrees of freedom of the CFT are not strictly speaking {\it particle} states).  In the original duality frame, the light ``tensionless string" degrees of freedom arise with $\ell\to 0$ from wrapped D3-branes (with appropriate worldvolume flux to cancel the contribution of $cg_s$ to the brane tension at the quantized values $c_j=(2\pi)^2j$).

Of order $10^3$ to $10^4$ circuits can fit inside the compactification, satisfying the back reaction constraint delineated in eqn (3.42) of \cite{McAllister:2008hb}\ by using the freedom to obtain somewhat large volume while maintaining high moduli stabilizing barriers using for example the methods of the large volume scenario \cite{LV}\ as explained in \S4.4.1\ of \cite{McAllister:2008hb}.

This construction corresponds to a linear potential $V(\phi)=\mu^3\phi$ (modulated by instanton-generated sinusoidal corrections), a simple generalization of the $m^2\phi^2$ model analyzed above in \S4.  This potential is slightly flatter but leads to similar conditions on its parameters, shown in figure \ref{fig:consistwindow}.

Now for a potential $V(\phi)=\mu^{4-\alpha}\phi^\alpha$, from our solution above we have
\beq\label{phideltratio} {\phi\over\Delta}\sim 10^{-15}g^{-25/6}(2\pi)^{-2} \left({M_P\over\phi}\right)^{7\alpha/12-5/6} \left({M_P\over\mu}\right)^{7/3-7\alpha/12} \ .\eeq
%
For $m^2\phi^2$, i.e. $\alpha=2$, this becomes (using (\ref{Ne}))
\beq\label{Ncycm} {\phi\over\Delta}\sim {10^{-18}\over (2\pi)^3 g^{9/2} N_{e}^{1/2}}\left({M_P \over m}\right)^{3/2}\ . \eeq
The number of e-foldings is
\beq\label{Negen} N_e={10^{-6}\over{g^{2/3} (2\pi)^{2}}}\left({\phi \over\mu}\right)^{4/3-\alpha/3}
\ , \eeq
which reproduces the result (\ref{Ne}) listed above for the case $\alpha=2$.

%
%

The non-Gaussianity constraint $\tilde m\lesssim 10H $ corresponds to (for $\alpha<4$)
\beq\label{NGparam}\left(\frac{M_{P}}{\mu}\right)^{4-\alpha} \frac{ 10^{-22+\alpha } N_e^{-1-\frac{\alpha }{2}}}{g^2 (2\pi)^6}\lesssim 1\ .\eeq
For $\alpha=2$ this becomes
\beq g {m\over M_P}\gtrsim {10^{-10}\over{N_e (2\pi)^{3}}} \ ,\eeq
and for $\alpha=1$ it is
\beq\label{NGmu} {\mu\over M_P}\gtrsim(2 \pi)^{-2} g^{-2/3}10^{-7} N_e^{-1/2}\ . \eeq
%

%
%

This corresponds to a constraint on the number of production events
\beq
N_{events}\sim\frac{\phi}{\Delta}\lesssim (2 \pi)^{3/2} 10^{-3} N_{e} g^{-3}\
\eeq
for any $\alpha$.
%
%
%
%
%
Similarly, one can derive a lower bound on $N_{events}$ from our slow roll conditions.  For a wide range of parameters, the most stringent condition comes from (\ref{eta2}).  For $\alpha<4$, the constraint is
\beq N_{events} \gg (2 \pi)^{3/2} 10^{-9/2}  N_e g^{-3} \ .\eeq
Therefore, for any $\alpha<4$ one has a window of $\sim10^{3/2}$ between the minimum and maximum values allowed.

Altogether, we find that the structure required for trapped inflation arises in the directions with monodromy in string compactifications, within a different regime of the potential and field range from that considered in modeling chaotic inflation in \cite{Silverstein:2008sg,McAllister:2008hb}.  The ingredients required for trapped inflation generally introduce more back reaction than occurs in the corresponding single-field chaotic inflation model, but do fit into a reasonable subset of the known constructions.

\section{Discussion}

One of the satisfying recent developments in inflationary theory has been a more systematic classification of inflationary mechanisms.  An inflationary mechanism can be characterized by its number of degrees of freedom -- single field versus multiple field (a feature correlated with $f_{\rm NL}^{\rm local}$ and isocurvature effects), the sound speed of its perturbations (correlated with $f_{\rm NL}^{\rm equilateral}$), and the field range of its inflaton (correlated with the gravity wave signature $r$).  The present mechanism involves multiple fields (including the $\chi$s), but it behaves like some single field models in its prediction for large $f_{\rm NL}^{\rm equilateral}$.\footnote{See e.g. \cite{Langlois:2008qf}\ for an analysis of possible shapes of the non-Gaussianity arising from multifield models with nontrivial kinetic terms.}


Although its signatures are somewhat similar to its strong-coupling analogue \cite{Alishahiha:2004eh}, we have seen that trapped inflation fits concretely into previously studied string compactifications; it is fair to say that the mechanism \cite{Silverstein:2003hf,Alishahiha:2004eh}\ lacks a known clean top-down embedding (in the small subset of string compactifications yet studied).  It would be interesting to find a compactification that interpolates between the two cases by varying the number of light degrees of freedom (and hence the 't Hooft coupling).

The calculations in this paper required somewhat novel techniques for treating the effective dynamics of $\phi$ resulting from the production of the sets of (temporarily) light $\chi$ particles.
There are several ways in which our analysis could be extended.  In particular, it would be useful to develop more precise analytical tools to treat the perturbations.


\section*{Acknowledgments}
We thank Tomas Rube for early collaboration.
We thank N. Arkani-Hamed, L. Kofman, A. Linde, X. Liu, L. McAllister, A. Westphal, and M. Zaldarriaga for useful discussions.
The research of D.G., B.H., and E.S. is supported by NSF grant
PHY-0244728, by the DOE under contract DE-AC03-76SF00515, and by BSF
and FQXi grants.  D.G. is also supported by a Mellam Family Graduate Fellowship and a NSERC Fellowship.  B. H. is also supported by a William K. Bowes Jr. Stanford Graduate Fellowship.  The research of LS is supported in part by the National Science Foundation under Grant No. PHY-0503584.


\begingroup\raggedright\endgroup

\end{document}